\documentclass[secnumarabic, amssymb, nobibnotes, aps, prd, twocolumn]{revtex4-1}

\usepackage{graphicx, color}
\usepackage{bm}

\begin{document}
\title{A feasibility study of the measurement of Higgs pair creation \\at a Photon Linear Collider}
\author{Shin-ichi Kawada$^1$}
\email{s-kawada@huhep.org}
\author{Nozomi Maeda$^1$}
\author{Tohru Takahashi$^1$}
\author{Katsumasa Ikematsu$^2$}
\author{Keisuke Fujii$^3$}
\author{Yoshimasa Kurihara$^3$}
\author{Koji Tsumura$^4$}
\author{Daisuke Harada$^5$}
\author{Shinya Kanemura$^6$}
\affiliation{$^1$Graduate School of Advanced Sciences of Matter, Hiroshima University, 1-3-1, Kagamiyama, Higashi-Hiroshima, 739-8530, Japan}
\affiliation{$^2$Department f\"{u}r Physik, Universit\"{a}t Siegen, D-57068, Siegen, Germany}
\affiliation{$^3$High Energy Accelerator Research Organization (KEK), 1-1, Oho, Tsukuba, Ibaraki, 305-0801, Japan}
\affiliation{$^4$Department of Physics, Graduate School of Science, Nagoya University, Furo-cho, Chikusa-ku, Nagoya, 464-8602, Japan}
\affiliation{$^5$Centre for High Energy Physics, Indian Institute of Science, Bangalore, 560012, India}
\affiliation{$^6$Department of Physics, University of Toyama, 3190 Gofuku, Toyama, 930-8555, Japan}
\date{\today}


\begin{abstract}
We studied the feasibility of the measurement of Higgs pair creation at a Photon Linear Collider (PLC).
From the sensitivity to the anomalous self-coupling of the Higgs boson, the optimum $\gamma \gamma$ collision energy was found to be around 270 GeV for a Higgs mass of 120 GeV/$c^2$.
We found that large backgrounds such as $\gamma \gamma \rightarrow W^+W^-, ZZ,$ and $b\bar{b}b\bar{b}$, can be suppressed if correct assignment of tracks to parent partons is achieved and Higgs pair events can be observed with a statistical significance of $\sim 5 \sigma$ by operating the PLC for 5 years.
\end{abstract}

\maketitle

\section{Introduction}

One of the most important events expected in particle physics in near future is unquestionably the discovery of the Higgs boson.
The data from the ATLAS and the CMS experiments at the LHC and the DZero and the CDF experiments at the Tevatron hint at the existence of a light Standard-Model-like Higgs boson in the mass range of 115 - 130 GeV/$c^2$ \cite{ATLAS, CMS, TEVATRON}.
If it is indeed the case, the discovery is expected to be declared within a year or so by the LHC experiments.

In the Standard Model, the Higgs boson is responsible for giving masses to both gauge bosons and matter fermions, via the gauge and Yukawa interactions, respectively, upon the spontaneous breaking of the electroweak symmetry.
However, unlike the gauge interaction, the mechanism of the spontaneous symmetry breaking and the Yukawa interaction have been left untested.
As a matter of fact, a Higgs doublet with its wine bottle potential, and its Yukawa coupling to each matter fermion in the Standard Model are mere assumptions other than being the minimal mechanism to generate the masses of gauge bosons and fermions.
In other words, we know essentially nothing but something must be condensed in the vacuum to give the masses of gauge bosons and fermions.
It is well known that the Standard Model cannot describe everything in the universe.
An example is the existence of the dark matter which occupies about one-fourth of the energy density in the universe.
The non-existence of anti-matter is another example.
Since the gauge sector of the Standard Model is well tested, it would be natural to expect that some hints of physics beyond the Standard Model could be obtained via precise measurements of the Higgs boson properties.

The LHC experiments are likely to discover the Standard-Model-like Higgs boson.
However, their precision is most probably not enough to reveal details of the discovered particle(s) due to high background environments of proton-proton collisions.
Thus, precise measurements of the Higgs boson properties by an electron-positron collider and its possible options are crucial to uncover its detailed properties which might go beyond the Standard Model.
The International Linear Collider (ILC) has potential to study the properties of the Higgs boson(s) such as coupling strengths to gauge bosons and matter fermions including the top quark with high precision, thereby opening up a window to physics beyond the Standard Model \cite{RDR2}.

\begin{figure}[!tb]
\centering
\includegraphics[scale=0.45]{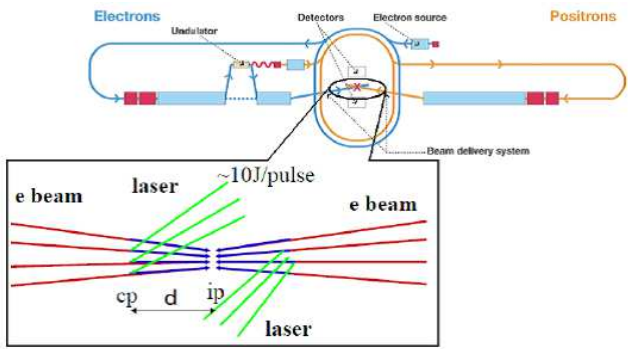}
\vspace*{-5pt}
\caption{A schematic of the PLC.
The positron beam of the ILC is replaced with an electron beam.}
\label{PLC}
\end{figure}

In addition to the $e^+e^-$ collisions, high energy photon-photon collisions are possible at the ILC by converting the electron beam to a photon beam by the inverse Compton scattering \cite{Compton}.
Physics and technical aspects of a Photon Linear Collider (PLC) as an option of the $e^+e^-$ Linear Collider are described, for instance, in ref. \cite{PLC2}.
A schematic of the PLC is shown in FIG. \ref{PLC}.
The Higgs boson properties such as its two-photon decay width and CP properties can be studied in high energy photon-photon interaction and thus the PLC plays complementary role to the $e^+e^-$ Linear Collider.
It should also be emphasized that the Higgs boson can be singly produced in the $s$-channel process so that the required electron beam energy is significantly lower than that for the $e^+e^-$ Linear Collider.

\begin{figure}[!tb]
\centering
\includegraphics[scale=0.22]{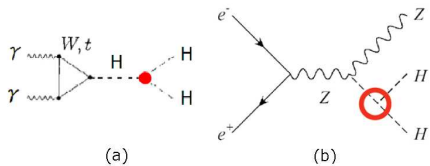}
\vspace*{-5pt}
\caption{Diagrams including the Higgs self-coupling, (a): for $\gamma \gamma \rightarrow HH$ and (b): for $e^+e^- \rightarrow ZHH$.
Higgs boson self-coupling occurs at red points.}
\label{AAHHeeZHH}
\end{figure}

One of the most important observables to be measured in the Higgs sector is its self-coupling, since it directly relates to the dynamics of the Higgs potential, {\it i.e.} the mechanism of the spontaneous symmetry breaking.
For example, a non-standard large deviation in the self-coupling can be direct evidence for strong first-order phase transition of the electroweak symmetry in the early universe \cite{kanemu}.
FIG. \ref{AAHHeeZHH} (a) and (b) show diagrams of processes which involve the self-coupling in $\gamma \gamma$ and $e^+e^-$ interactions.
Recently, a prospect for studying the self-coupling at the ILC was reported.
According to the study, the self-coupling is expected to be measured with precision of 57 \% with an integrated luminosity of 2 ab$^{-1}$ at $\sqrt{s} = 500$ GeV \cite{Tian}.

Measurements of the self-coupling at the PLC were discussed by several authors \cite{BJ,PLC,kanemura}.
It has been pointed out that contributions to the self-coupling to the cross-section of FIG. \ref{AAHHeeZHH} (a) and (b) are different and measurements in $e^+e^-$ and $\gamma \gamma$ interactions are complementary from physics point of view.
In addition, as a Higgs boson pair is directly produced in the $\gamma \gamma$ interaction, required beam energy is lower, 190 GeV as described later, than that for the $e^+e^-$ interaction.
This nature is important when considering energy update scenarios of the ILC.

In ref. \cite{BJ}, an order-of-magnitude estimation for background processes was presented.
However, the cross-section for the $W$ boson pair production is $10^6$ orders of higher than that for the Higgs boson pair production.
The backgrounds from $b\bar{b}b\bar{b}$, $b\bar{b}c\bar{c}$, and $ZZ$ production processes are also large and have the same final state as with the Higgs pairs for a low mass Higgs boson, which predominantly decays into $b\bar{b}$.
Given the situation, people had been skeptical about the feasibility of the detection of the Higgs pair process at the PLC.

In this work, we studied, for the first time, the Higgs boson pair creation at the PLC extensively with a parameter set of the PLC based on an $e^+e^-$ linear collider optimized for the light Higgs boson of 120 GeV/$c^2$ and the same detector simulation framework as used for the ILC physics analysis.
We report details of the analysis, issues and prospects for the measurement of the Higgs boson pairs at the PLC.

\section{Beam parameters}

\begin{figure}[!tb]
\centering
\includegraphics[scale=0.4]{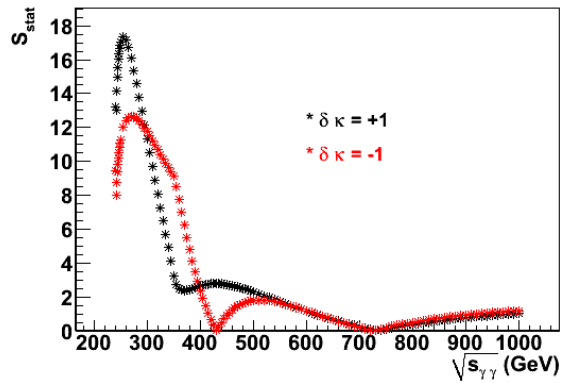}
\vspace*{-5pt}
\caption{Statistical sensitivity ($S_{\mathrm{stat}}$) as a function of $\gamma \gamma$ collision energy.
Black and red dots show the $\delta \kappa = +1$ and $\delta \kappa = -1$ cases.}
\label{sensitivity}
\end{figure}

In order to choose parameters for the PLC, we calculated the statistical sensitivity $S_{\mathrm{stat}}$ defined as
\begin{equation}
S_{\mathrm{stat}} = \frac{|N(\delta \kappa)-N_{\mathrm{SM}}|}{\sqrt{N_{\mathrm{obs}}}} 
= \frac{L|\eta \sigma (\delta \kappa) - \eta \sigma _{\mathrm{SM}}|}{\sqrt{L(\eta \sigma (\delta \kappa) + \eta _{\mathrm{BG}}\sigma _{\mathrm{BG}})}},
\end{equation}
where, $\delta \kappa$ is the deviation of the self-coupling constant from the Standard Model.
The constant of Higgs self-coupling $\lambda$ can be expressed as $\lambda = \lambda ^{\mathrm{SM}}(1 + \delta \kappa)$, where $\lambda ^{\mathrm{SM}}$ is the Higgs self-coupling constant in the Standard Model.
$N(\delta \kappa)$ and $N_{\mathrm{SM}}$ are the expected number of events as a function of $\delta \kappa$ and that expected from the Standard Model.
$\sigma (\delta \kappa)$ and $\sigma _{\mathrm{SM}}$ are the cross-section of the Higgs boson production as a function of $\delta \kappa$ and that of the Standard Model, while $L$,  $\eta$, $\eta _{\mathrm{BG}}$, and $\sigma _{\mathrm{BG}}$ are the integrated luminosity, the detection efficiency for the signal, the detection efficiency for backgrounds, and the cross-section of background processes, respectively.
For $\eta = 1$ and $\eta _{\mathrm{BG}} = 0$, $S_{\mathrm{stat}}$ is written as
\begin{equation}
S_{\mathrm{stat}} = \sqrt{L} \frac{|\sigma (\delta \kappa) - \sigma _{\mathrm{SM}}|}{\sqrt{\sigma (\delta \kappa)}}.
\end{equation}
FIG. \ref{sensitivity} plots $S_{\mathrm{stat}}$ as a function of the center of mass energy of the $\gamma \gamma$ collision (denoted $\sqrt{s_{\gamma \gamma}}$ hereafter) for the Higgs boson mass of 120 GeV$/c^2$ with the $\gamma \gamma$ integrated luminosity of 1000 fb$^{-1}$.
The cross-section of the signal was calculated according to the formula described in ref. \cite{PLC} for the case of $\delta \kappa = +1$ and $\delta \kappa = -1$ as indicated in FIG. \ref{sensitivity}.
From this result, we found the optimum energy to be $\sqrt{s_{\gamma \gamma}} \approx$ 270 GeV.

\begin{figure}[!tb]
\centering
\includegraphics[scale=0.42]{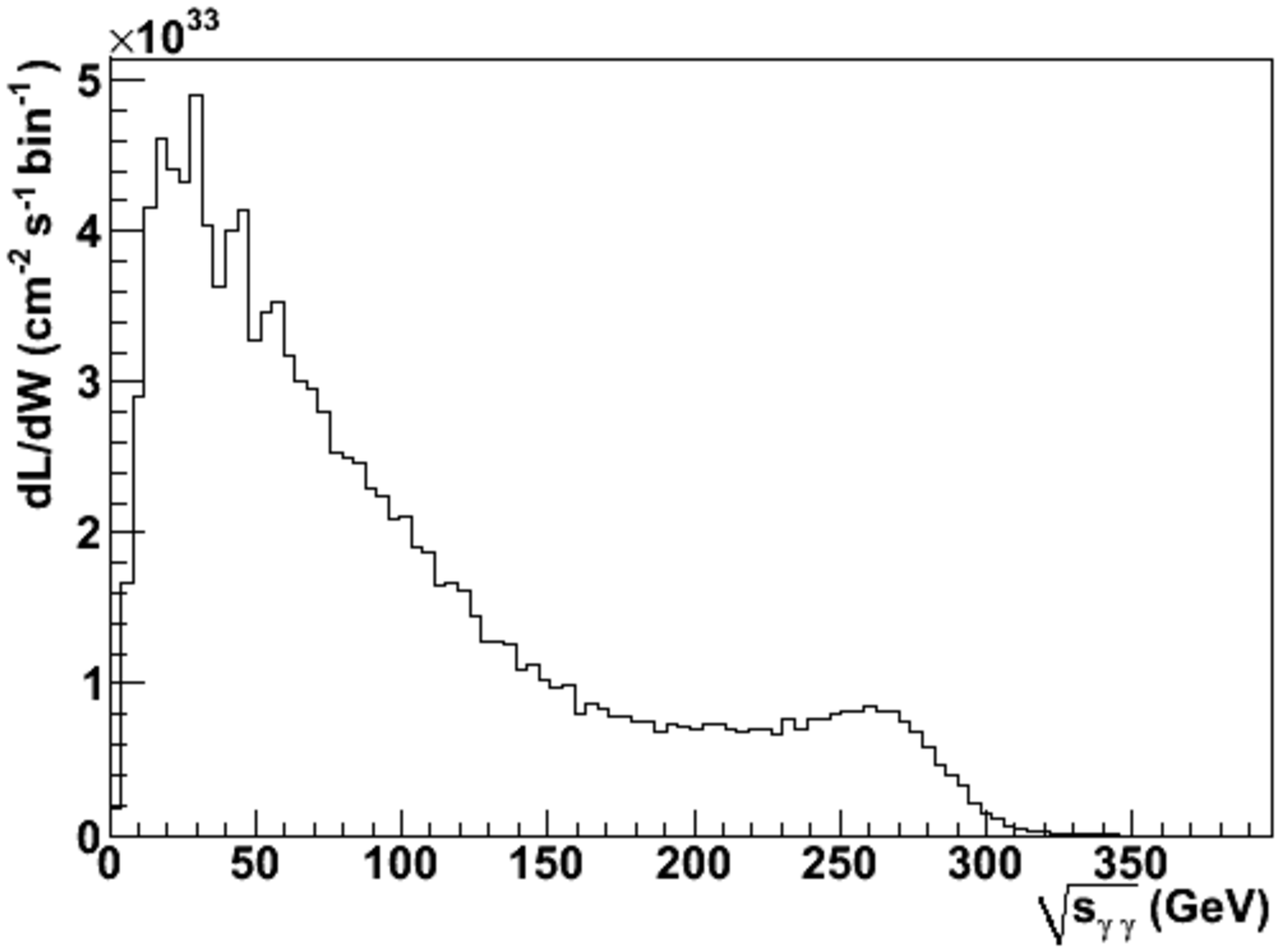}
\vspace*{-5pt}
\caption{Luminosity distribution generated by CAIN.
Input parameters are shown in TABLE \ref{CAIN}.}
\label{luminosity}
\end{figure}

The parameters for the electron and the laser beams are summarized in TABLE \ref{CAIN}.
It was designed to maximize $\gamma \gamma$ luminosity at $\sqrt{s_{\gamma \gamma}} \approx 270$ GeV based on the TESLA optimistic parameters \cite{param}.
The wavelength of the laser was chosen to be 1054 nm, which is a typical wavelength for solid state lasers.
The electron beam energy was chosen to maximize $\gamma \gamma$ luminosity around 270 GeV, while keeping the electron beam emittance and the $\beta$ functions at the interaction point the same as the TESLA parameters.
The luminosity distribution was simulated by CAIN \cite{CAIN}, as shown in FIG. \ref{luminosity}.
The $\gamma \gamma$ luminosity in the high energy region ($\sqrt{s_{\gamma \gamma}} > 0.8 \sqrt{s_{\gamma \gamma}^{\mathrm{max}}}$) was calculated to be $1.2\times 10^{34} \mathrm{cm}^{-2}\mathrm{s}^{-1}$.

\begin{table}[!tb]
\centering
\caption{The parameters of electron and laser beams based on TESLA optimistic parameters.
The polarization of electron beam was assumed to be 100 {\%}.}
\begin{ruledtabular}
\begin{tabular}{ccc}
parameter & unit &  \\ \hline
electron beam energy & $E_e$ [GeV] & 190 \\
\# of electrons / bunch & $N\times 10^{10}$ & 2 \\
longitudinal beam size & $\sigma _z$ [mm] & 0.35 \\
transverse emittance & $\gamma \varepsilon _{x/y}$ [$10^{-6}$m$\cdot$rad] & 2.5/0.03 \\
$\beta$ function @ IP & $\beta _{x/y}$ [mm] & 1.5/0.3 \\
transverse beam size & $\sigma _{x/y}$ [nm] & 100/5.7 \\
laser wavelength & $\lambda _L$ [nm] & 1054 \\
laser pulse energy & [J] & 10 \\
$x = 4\omega E_e/m_e^2$ &  & 3.42 \\
\end{tabular}
\end{ruledtabular}
\label{CAIN}
\end{table}

\section{Signal and Backgrounds}

\begin{figure}[!tb]
\centering
\includegraphics[scale = 0.41]{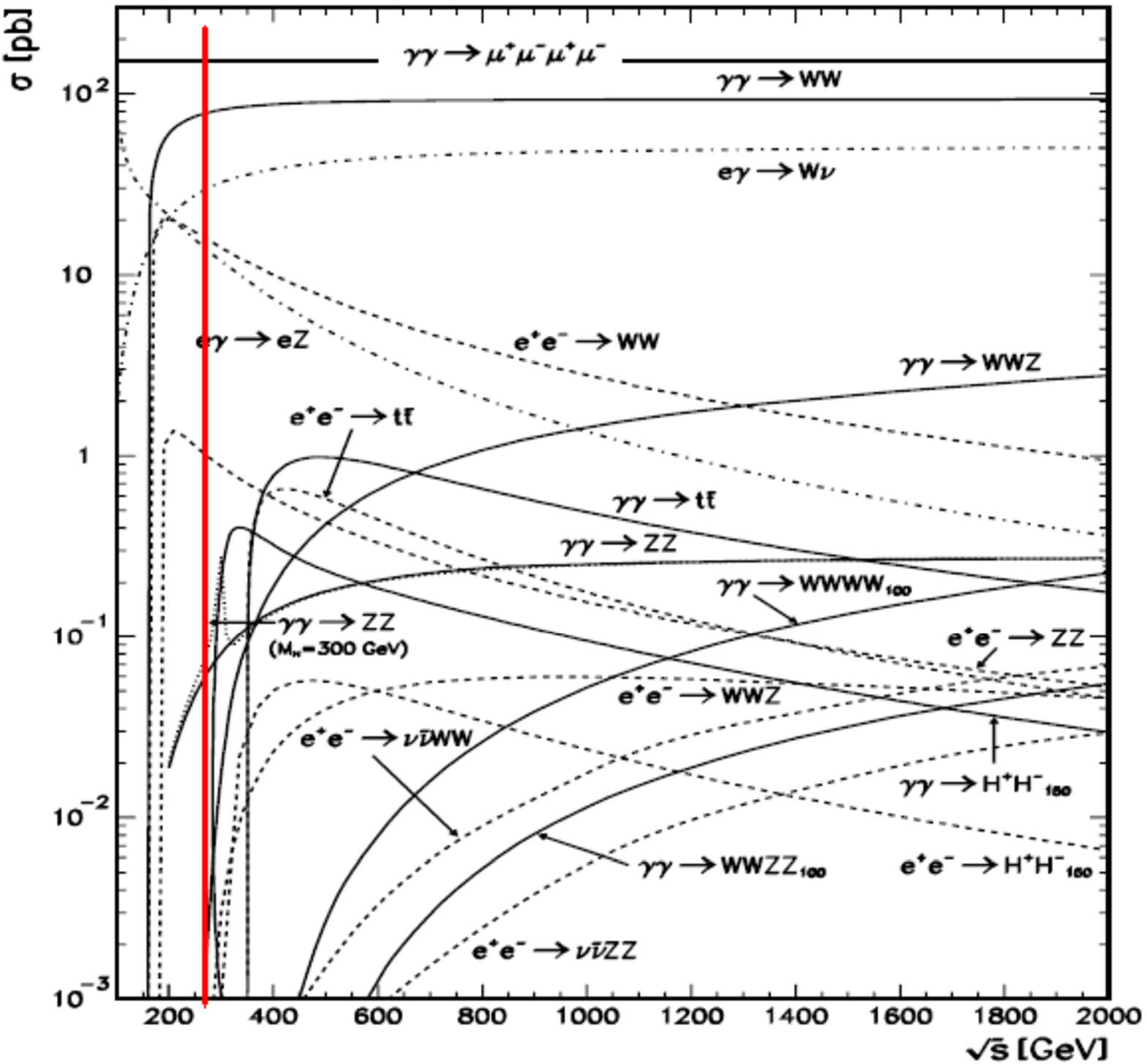}
\vspace*{-5pt}
\caption{The cross-sections of various Standard Model processes as a function of collision energy.
Solid lines show the $\gamma \gamma$ collision case. 
The red line shows the optimum energy, 270 GeV.}
\label{cross-section}
\end{figure}

FIG. \ref{cross-section} shows the cross-sections for various processes of $\gamma \gamma$ and $e^+e^-$ collisions as a function of the center of mass energy.
The figure indicates that the $\gamma \gamma \rightarrow WW$ and $\gamma \gamma \rightarrow ZZ$ processes will be the main backgrounds at $\sqrt{s_{\gamma \gamma}} =$ 270 GeV because the total cross-sections are about 90 pb and 60 fb, respectively, far exceeding that of $\gamma \gamma \rightarrow HH$, which was calculated to be 0.19 fb.
It should be noted that $\sqrt{s_{\gamma \gamma}} =$ 270 GeV is below the threshold of the $\gamma \gamma \rightarrow t\bar{t}$ process so that it is not necessary to be considered as a background source.

TABLE \ref{B.R.} shows the branching ratios of the Standard Model Higgs boson with a mass of 120 GeV/$c^2$ \cite{maeda}.
Since the main decay mode of the 120 GeV/$c^2$ Higgs boson is $H \rightarrow b\bar{b}$, we concentrated on the case where both Higgs bosons decay into $b\bar{b}$ in this analysis.
This implies that the $\gamma \gamma \rightarrow b\bar{b}b\bar{b}$ process must also be considered as a possible background process.

\begin{table}[!tb]
\centering
\caption{Branching ratios of the Standard Model Higgs boson with a mass of 120 GeV/$c^2$.}
\begin{ruledtabular}
\begin{tabular}{p{10pt}rlp{10pt}}
& decay mode & branching ratio & \\  \hline
\rule[5pt]{0pt}{5pt}
& $H \rightarrow b \bar{b}$ & 0.68 & \\
& $H \rightarrow WW^*$ & 0.13 & \\
& $H \rightarrow gg$ & 0.071 & \\
& $H \rightarrow \tau \tau$ & 0.069 & \\
& $H \rightarrow c \bar{c}$ & 0.030 & \\
& $H \rightarrow ZZ^*$ & 0.015 & \\
& $H \rightarrow \gamma \gamma$ & 0.0022 & \\
& $H \rightarrow \gamma Z$ & 0.0011 & \\
& $H \rightarrow s \bar{s}$ & 0.00051 & \\
& $H \rightarrow \mu \mu$ & 0.00024 & \\
\end{tabular}
\end{ruledtabular}
\label{B.R.}
\end{table}

The numbers of events expected for the signal and the backgrounds were calculated from the $\gamma \gamma$ cross-sections by convoluting them with the luminosity distribution, as
\begin{equation}
N_{\mathrm{events}} = \int \sigma (W_{\gamma \gamma}) \frac{dL}{dW_{\gamma \gamma}} dW_{\gamma \gamma}.
\end{equation}
We used the formula in refs. \cite{PLC,kanemura} for the calculation of $\gamma \gamma \rightarrow HH$, HELAS \cite{HELAS} for $\gamma \gamma \rightarrow WW$, gamgamZZ-code \cite{gamgamZZ-1, gamgamZZ-2} and HELAS for $\gamma \gamma \rightarrow ZZ$, and GRACE \cite{GRACE} for $\gamma \gamma \rightarrow b\bar{b}b\bar{b}$.
The numerical integration and subsequent event generation was performed by BASES/SPRING \cite{BS}.
With this calculation, we expect 16 events/year for $\gamma \gamma \rightarrow HH$, $1.462\times 10^7$ events/year for $\gamma \gamma \rightarrow WW$, and $1.187\times 10^4$ events/year for $\gamma \gamma \rightarrow ZZ$. 
For $\gamma \gamma \rightarrow b\bar{b}b\bar{b}$, $5.194 \times 10^4$ events/year is estimated for events with $b\bar{b}$ mass grater than 15 GeV/$c^2$.

\section{Simulation and Analysis}

JLC Study Framework (JSF) \cite{JLC,JSF} was used as our simulation framework in this study.
Pythia6.4 \cite{Pythia} was used for parton shower evolution and subsequent hadronaization.
For the detector simulation, a fast simulator, QuickSim \cite{JSF}, was used instead of a full detector simulation in order to process the huge background samples.

QuickSim is, however, fairly detailed and realistic: it smears track parameters with their correlations, vertex detector hits according to given resolution and multiple scattering.
It simulates calorimeter signals to individual cells so as to take property into account their possible overlapping.
The calorimeter signals are then clustered and matched to charged tracks, if any, to form particle-flow-like objects to archive best attainable jet energy resolution (see ref. \cite{ttH} for more details).

\begin{table}[!tb]
\centering
\caption{The detector parameters.
$p, p_T$ and $E$ are measured in units GeV.
The angle $\theta$ is measured from the beam axis.}
\begin{ruledtabular}
\begin{tabular}{cc}
Detector & Resolution \\  \hline
Vertex detector & $\sigma _b = 7.0 \oplus (20.0/p\sin ^{3/2}\theta)$ $\mu$m \\
Drift chamber & $\sigma _{p_T}/p_T = 1.1 \times 10^{-4}p_T \oplus 0.1{\%}$ \\
ECAL & $\sigma _E/E = 15{\%}/\sqrt{E} \oplus 1{\%}$ \\
HCAL & $\sigma _E/E = 40{\%}/\sqrt{E} \oplus 2{\%}$
\end{tabular}
\end{ruledtabular}
\label{detparam}
\end{table}

The detector parameters are summarized in TABLE \ref{detparam}.
In the simulation, we assumed a dead cone of a half angle of $7.6^{\circ}$ in the forward/backward region of the detector to house the laser optics, the beam pipes and the masking system \cite{masking}.

We generated $5 \times 10^4$ Monte-Carlo events for $\gamma \gamma \rightarrow HH$, $7.5 \times 10^7$ for $\gamma \gamma \rightarrow WW$, $1 \times 10^6$ for $\gamma \gamma \rightarrow ZZ$, and $1 \times 10^6$ for $\gamma \gamma \rightarrow b\bar{b}b\bar{b}$, respectively, which are statistically sufficient to assess the feasibility of $\gamma \gamma \rightarrow HH$ measurement against the large number of background events.

\subsection{Event Selection}

First, we applied the forced 4-jet clustering to each event in which the clustering algorithm was applied to each event by changing the clustering parameter until the event is categorized as a 4-jet event.
We used the JADE clustering \cite{JADE} as the clustering algorithm.

Using four-momenta of reconstructed jets, $\chi ^2_i$s ($i = H, W, Z, b\bar{b}$) were calculated for possible jet combinations as
\begin{equation}
\chi _i^2 = \mathrm{min}\left[ \frac{(M_1 - M_i)^2}{\sigma _{2ji}^2} + \frac{(M_2 - M_i)^2}{\sigma _{2ji}^2}\right],
\end{equation}
where, $M_1$ and $M_2$ are invariant masses of two jets.
$M_i$ ($i = H, W, Z, b\bar{b}$) are the masses of the Higgs boson, the $W$ boson, the $Z$ boson, and the invariant mass of $b\bar{b}$ (10 GeV/$c^2$), respectively.
$\sigma _{2ji}$ ($i = H, W, Z, b\bar{b}$) are their corresponding mass resolutions, and are chosen to be $\sigma _{2jH} = 8$ GeV and $\sigma _{2jW} = \sigma _{2jZ} = \sigma _{2jb\bar{b}} = 6$ GeV, respectively.
The ``min[ \ ]" stands for the operation to choose the minimum out of all the jet combinations.

\begin{figure}[!t]
\centering
\includegraphics[scale = 0.3]{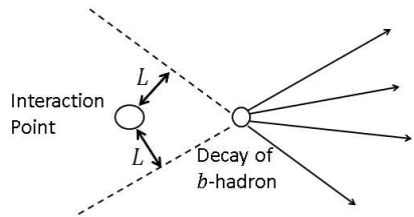}
\vspace*{-5pt}
\caption{The concept of ``nsig" method.
A $b$-hadron is generated at ``Interaction Point" and decays at the circle indicated as ``Decay of $b$-hadron."
Arrows represent particle tracks.
Dotted lines are some of the extrapolated particle tracks towards the interaction point.}
\label{nsig}
\end{figure}

In order to discriminate $b$-quarks, we used the ``nsig" method for the $b$-tagging in this study.
FIG. \ref{nsig} illustrates the concept of the ``nsig" method.
For each track in a reconstructed jet, $N_{\mathrm{sig}} = L/\sigma _L$ was calculated, where $L$ is the distance of closest approach to the interaction point of the track in the plane perpendicular to the beam and $\sigma _L$ is its resolution.
Then $N_{\mathrm{offv}}(a)$, the number of tracks which have $N_{\mathrm{sig}} > a$, is calculated for each jet as a function of $a$.

Before optimizing event selection criteria, we applied the pre-selection to reduce the number of background events to a level applicable to the Neural Network analysis.
The criteria for the pre-selection are
\begin{itemize}
\item $N_{\mathrm{jet}}(N_{\mathrm{offv}}(3.0) \ge 1) \ge 3$,
\item $N_{\mathrm{jet}}(N_{\mathrm{offv}}(3.0) \ge 2) \ge 2$,
\item $\beta _{2j}> 0.05$,
\item $|\cos \theta _{2j}| < 0.99$,
\end{itemize}
where, $N_{\mathrm{jet}}(N_{\mathrm{offv}}(b) \ge c)$ is the number of jets for which $N_{\mathrm{offv}}(b)$ is greater than or equal to $c$.
$\beta _{2j}$ is the speed of reconstructed 2-jet which has the least $\chi ^2_i$  and $\theta _{2j}$ is the angle of reconstructed 2-jet system with respect to the beam axis.

After the pre-selection, we applied the Neural Network analysis to optimize the selection criteria.
It is a three-layer network with a single output.
JETNET \cite{NN} was used to train the Neural Network system which employed the back propagation for the weight optimization.

\begin{figure*}[!tb]
\centering
\includegraphics[scale = 0.285]{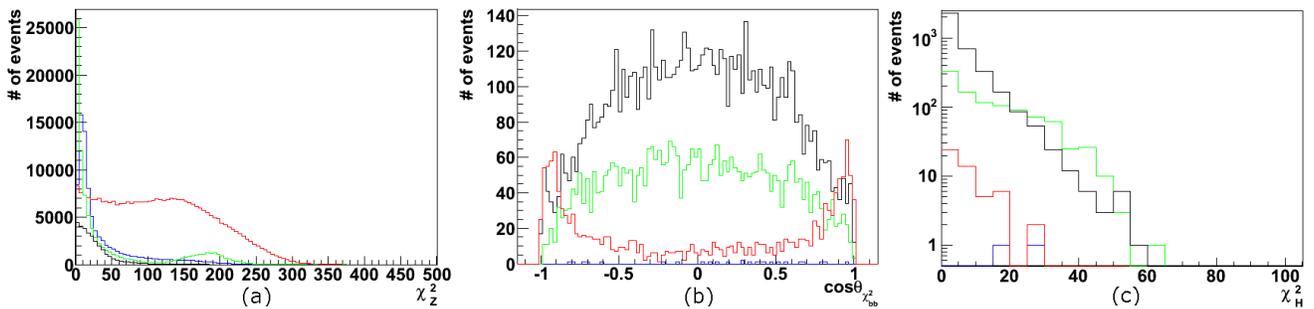}
\vspace*{-5pt}
\caption{Typical distributions of Neural Network input variables.
(a): $\chi ^2_Z$ after the pre-selection.
(b): $\cos \theta _{\chi ^2_{b\bar{b}}}$ after the $W$ filter.
(c): $\chi ^2_H$ after the $b\bar{b}$ filter.
Black, blue, green, and red histograms show the $\gamma \gamma \rightarrow HH$ (signal), $\gamma \gamma \rightarrow WW$, $\gamma \gamma \rightarrow ZZ$, and $\gamma \gamma \rightarrow b\bar{b}b\bar{b}$ events, respectively.}
\label{input_JADE}
\end{figure*}

For $\gamma \gamma \rightarrow WW$ events, inputs to the Neural Network are $\chi ^2_H$, $\chi ^2_Z$, the visible energy, $N_{\mathrm{jet}}(N_{\mathrm{offv}}(3.5) \ge 1)$, $N_{\mathrm{jet}}(N_{\mathrm{offv}}(3.5) \ge 2)$, the longitudinal momentum, the transverse momentum, the number of tracks, and $\mathrm{Y_{cut}}$ of jet clustering.
29958 signal events and 83777 background events were used for Neural Network training with the number of intermediate layers of 18.
FIG. \ref{input_JADE} (a) shows the typical distribution of $\chi ^2_Z$.

Neural Network inputs for $\gamma \gamma \rightarrow b\bar{b}b\bar{b}$ analysis are $\chi ^2_H$, $\chi ^2_{b\bar{b}}$, $\cos \theta _{\chi ^2_H}$, $\cos \theta _{\chi ^2_{b\bar{b}}}$, the visible energy, the number of tracks, $\mathrm{Y_{cut}}$ of jet clustering, thrust \cite[pp. 284]{Collider}, sphericity \cite[pp .281]{Collider}, $Y$ value \cite[pp .282]{Collider}, $\cos \theta _j$, and the largest $|\cos \theta _j|$ of the event, where $\theta _{\chi ^2_H (\chi ^2_{b\bar{b}})}$ and $\theta _j$ are the angle of $H (b\bar{b})$ system and of each jet, with respect to the beam axis.
FIG. \ref{input_JADE} (b) shows the distribution of $\cos \theta _{\chi ^2_{b\bar{b}}}$ after the $W$ filter.
7756 and 1409 events for the signal and background, respectively, were used with the number of intermediate layers of 34.

For the $\gamma \gamma \rightarrow ZZ$ events, we used $\chi ^2_H$, $\chi ^2_W$, $\chi ^2_Z$, the visible energy, the number of tracks, the longitudinal momentum, the energies of the 2-jet systems, $N_{\mathrm{jet}}(N_{\mathrm{offv}}(3.5) \ge 1)$, and $N_{\mathrm{jet}}(N_{\mathrm{offv}}(3.5) \ge 2)$ as Neural Network inputs with 4536 signal and 1189 background events for the training with 20 intermediate layers.
FIG. \ref{input_JADE} (c) shows the typical distribution of $\chi ^2_H$.

The Neural Network was trained to maximize statistical significance $\Sigma$ defined as
\begin{equation}
\Sigma \equiv \frac{N_{\mathrm{signal}}}{\sqrt{N_{\mathrm{signal}} + N_{\mathrm{BG}}}},
\end{equation}
where, $N_{\mathrm{signal}}$ and $N_{\mathrm{BG}}$ are the numbers of remaining signal and background events, respectively.
To reduce possible systematic effects from the training of the Neural Network analysis, the performance of the Neural Network was evaluated by applying the results of the training (weight files) to events generated separately from the training samples.
In order to reduce the effect of the statistics of the event samples, we prepared the same number of events for the test sample for each training sample.
TABLE \ref{Cutsummary} shows the summary of event selection with JADE clustering.
From TABLE \ref{Cutsummary}, the statistical significance with the JADE clustering $\Sigma _{\mathrm{JADE}}$ was calculated to be
\begin{equation}
\Sigma _{\mathrm{JADE}} = 0.922 ^{+0.045}_{-0.067} \sigma.
\end{equation}

\begin{table*}[!tb]
\centering
\caption{Cut statistics with JADE clustering.
The numbers in the table are the expected numbers of surviving events expected in 5 years.
The error on each number is from statistics of the Monte-Carlo study.}
\begin{ruledtabular}
\begin{tabular}{rcccc}
 & $\gamma \gamma \rightarrow HH$ & $\gamma \gamma \rightarrow WW$ & $\gamma \gamma \rightarrow ZZ$ & $\gamma \gamma \rightarrow b\bar{b}b\bar{b}$ \\ \hline
expected events & 80 & $7.31\times 10^7$ & 59350 & 259700 \\
pre-selection & 47.72$\pm$0.28 & 81312$\pm$282 & 5172$\pm$18 & 80002$\pm$144 \\
applying $W$ filter & 12.27$\pm$0.14 & 24.4$\pm$4.9 & 231.6$\pm$3.7 & 378.1$\pm$9.9 \\
applying $b\bar{b}$ filter & 5.867$\pm$0.097 & 1.95$^{+2.6}_{-0.61}$ & 59.3$\pm$1.9 & 13.2$\pm$1.9 \\
applying $Z$ filter & 3.766$\pm$0.078 & 0$^{+1.8}_{-0}$ & 5.40$\pm$0.57 & 7.5$\pm$1.4 \\
\end{tabular}
\end{ruledtabular}
\label{Cutsummary}
\end{table*}

\subsection{Event Selection with an Ideal Clustering}

\begin{figure*}[!tb]
\centering
\includegraphics[scale = 0.285]{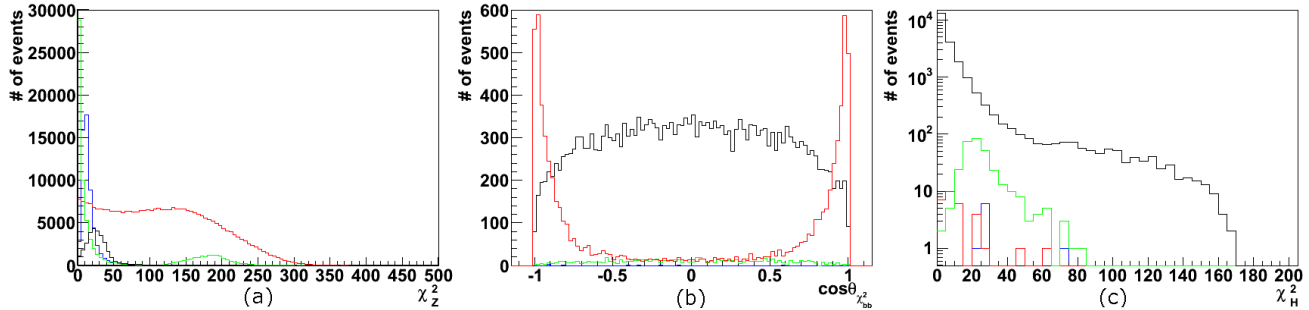}
\vspace*{-5pt}
\caption{Typical distributions of input variables in the case of the ideal jet clustering.
(a): $\chi ^2_Z$ after the pre-selection.
(b): $\cos \theta _{\chi ^2_{b\bar{b}}}$ after the $W$ filter.
(c): $\chi ^2_H$ after the $b\bar{b}$ filter.
Black, blue, green, and red histograms show the $\gamma \gamma \rightarrow HH$, $\gamma \gamma \rightarrow WW$, $\gamma \gamma \rightarrow ZZ$, and $\gamma \gamma \rightarrow b\bar{b}b\bar{b}$ events, respectively.}
\label{input_color}
\end{figure*}

The result in the previous section indicated that it is necessary to improve the performance of event selection.
In order to evaluate the effect of the jet clustering, we applied an ``ideal jet clustering'' to the $\gamma \gamma \rightarrow HH$, $\gamma \gamma \rightarrow WW$, and $\gamma \gamma \rightarrow ZZ$ events where each track is assigned to its parent ($H$, $W$, or $Z$) by color information obtained from the event generators.
The ``ideal jet clustering" was not applied to the $\gamma \gamma \rightarrow b\bar{b}b\bar{b}$ events since the color singlet combinations were non-trivial for this process.
Input variables to the Neural Network and the number of intermediate layers are the same as the JADE clustering case.
As with the previous analysis, the pre-selection was applied and events survived the selection cuts were used for the Neural Network analysis.
The number of signal/background events used for the Neural Network training were, 29152/57058 for $\gamma \gamma \rightarrow WW$, 24305/6349 for $\gamma \gamma \rightarrow b\bar{b}b\bar{b}$, and 22823/291 for $\gamma \gamma \rightarrow ZZ$, respectively.
FIG. \ref{input_color} (a), (b), and (c) show the typical distributions of $\chi ^2_Z$ after the pre-selection, $\cos \theta _{\chi ^2_{b\bar{b}}}$ after the $W$ filter, and $\chi ^2_H$ after the $b\bar{b}$ filter, respectively.
We again applied the results of the Neural Network training to the event samples which are statistically independent of the training samples.
TABLE \ref{Cutsummary_cheat} shows the summary of the event selection with the ideal jet clustering.
From TABLE \ref{Cutsummary_cheat}, the significance $\Sigma _{\mathrm{ideal}}$ was calculated to be
\begin{equation}
\Sigma _{\mathrm{ideal}} = 4.87 \pm 0.13 \sigma.
\end{equation}
This result indicates that $\gamma \gamma \rightarrow HH$ would be observed at $\sim 5 \sigma$ significance level with the integrated luminosity that corresponds to 5-year operation of the PLC, if the jet clustering performed perfectly.

\begin{table*}[!tb]
\centering
\caption{The similar table to TABLE \ref{Cutsummary}, but with the ideal jet clustering.}
\begin{ruledtabular}
\begin{tabular}{rcccc}
 & $\gamma \gamma \rightarrow HH$ & $\gamma \gamma \rightarrow WW$ & $\gamma \gamma \rightarrow ZZ$ & $\gamma \gamma \rightarrow b\bar{b}b\bar{b}$ \\ \hline
expected events & 80 & $7.31\times 10^7$ & 59350 & 259700 \\
pre-selection & 46.64$\pm$0.27 & 55836$\pm$233 & 4172$\pm$16 & 77778$\pm$142 \\
applying $W$ filter & 40.13$\pm$0.25 & 7.8$\pm$2.8 & 46.3$\pm$1.7 & 1826$\pm$22 \\
applying $b\bar{b}$ filter & 36.03$\pm$0.24 & 7.8$\pm$2.8 & 18.5$\pm$1.0 & 7.8$\pm$1.4 \\
applying $Z$ filter & 34.68$\pm$0.24 & 4.9$\pm$2.2 & 5.22$\pm$0.56 & 6.0$\pm$1.2 \\
\end{tabular}
\end{ruledtabular}
\label{Cutsummary_cheat}
\end{table*}

\section{Summary}

We studied the feasibility of the measurement of Higgs pair creation at the PLC, which is a possible option of the ILC.
The optimum center of mass energy of the $\gamma \gamma$ collision was found to be around 270 GeV for the Higgs boson with a mass of 120 GeV/$c^2$.

We found that the $\gamma \gamma \rightarrow HH$ process can be observed with a statistical significance of about $5 \sigma$ for the integrated luminosity corresponding to 5 years of the PLC running against the background process which has $10^6$ times larger production cross-section ($\gamma \gamma \rightarrow WW$) than the signal and other backgrounds which have the same final state ($\gamma \gamma \rightarrow ZZ$, and $\gamma \gamma \rightarrow b\bar{b}b\bar{b}$), if each track could be successfully assigned to parent particles (or partons).

Our analysis showed, for the light Higgs boson, improvement of the jet clustering technique is crucial to discriminate the backgrounds by invariant mass information rather than to improve the $b$-quark tagging efficiency.
This fact is reasonable, because the $WW$ background turned out to be suppressed by a simple $b$-quark tagging scheme since the $W$ bosons do not decay into $b$-quark pairs, while the $ZZ$ and $b\bar{b}b\bar{b}$ backgrounds can only be suppressed by their mass differences.

For further improvements, vertex information from $b$-tagging analysis must be taken into account in jet clustering, thus they should be coherently developed.
Efforts in this direction are on-going as a part of the ILC physics study \cite{clustering,flavor} and significant improvement could be expected in near future.

This analysis shows possibility to measure the Higgs boson self-coupling at a lower beam energy than that of the $e^+e^-$ mode and is useful in considering energy upgrade scenarios of the ILC.

\begin{acknowledgments}

The authors would like to thank ILC physics working group \cite{ILCphys} for useful discussions, especially Y. Okada for useful theoretical inputs.
This work is supported in part by the Creative Scientific Research Grant No. 18GS0202 of the Japan Society for Promotions of Science (JSPS), the JSPS Core University Program, and the JSPS Grant-in-Aid for Science Research No. 22244031, and the JSPS Specially Promoted Research No. 23000002.

\end{acknowledgments}



\end{document}